\documentclass[groupedaddress,
  aip,
  pop,
  amsmath,amssymb,
  reprint,
]{revtex4-1}
\usepackage{graphicx,color}
\usepackage{amsmath}
\usepackage{natbib}
\usepackage{epsfig}
\begin{document}

\title{\color{blue} Researh Note: Stokes-Einstein relation in simple fluids revisited }

\author{Sergey Khrapak}
\affiliation{Institut f\"ur Materialphysik im Weltraum, Deutsches Zentrum f\"ur Luft- und Raumfahrt (DLR), 82234 We{\ss}ling, Germany;
Joint Institute for High Temperatures, Russian Academy of Sciences, 125412 Moscow, Russia}

\begin{abstract}
In this Research Note the Zwanzig's formulation of the Stokes-Einstein (SE) relation for simple atomistic fluids is re-examined. It is shown that the value of the coefficient in SE relation depends on the ratio of the transverse and longitudinal sound velocities. In some cases, this ratio can be directly related to the pair interaction potential operating in fluids and thus there can be a certain level of predictivity regarding the value of this coefficient. This Research Note provides some evidence in favour of this observation. In particular,  
analyzing the situation in several model systems such as one-component plasma, Yukawa, inverse-power-law, Lennard-Jones, and hard-sphere fluids, it is demonstrated that there are certain correlations between the interaction softness and the coefficient in SE relation. The SE coefficient is also re-evaluated for various liquid metals at the melting temperature, for which necessary data are available.

\end{abstract}

\date{\today}

\maketitle

The conventional Stokes-Einstein (SE) relation expresses the diffusion coefficient $D$ of a tracer macroscopic spherical (``Brownian'') particle of radius $R$ in terms of the temperature $T$ and shear viscosity $\eta$ coefficient of a medium it is immersed in. It reads
\begin{equation}\label{SE_1}
D=\frac{T}{c\pi \eta R},
\end{equation}   
where $c$ is a numerical coefficient: $c=6$ or $c=4$ corresponds to the ``stick'' or ``slip'' boundary condition at the sphere surface, respectively. When the size of the tracer sphere decreases and tends to the size of atoms or molecules of the medium, the actual size of the sphere in Eq.~(\ref{SE_1}) has to be replaced by the so-called hydrodynamic radius $R_{\rm H}$, which can depend on details of the interaction between the sphere and the atoms or molecules of the medium. 
Going further down to atomistic scales, when self-diffusion of atoms in simple pure fluids is considered, the SE relation takes the form~\cite{Gaskell1982,ZwanzigJCP1983,OhtoriJCP2018,KhrapakAIPAdv2018,CostigliolaJCP2019}
\begin{equation}\label{SE_2}
D\eta(\Delta/T)=\alpha,
\end{equation}
where $\Delta=\rho^{-1/3}$ is the mean interparticle separation, which now plays the role of the effective tracer sphere diameter, and $\rho$ is the density. Sometimes Eq.~(\ref{SE_2}) is written in terms of the Wigner-Seitz radius $a = (4\pi \rho/3)^{-1/3}\simeq 0.620\Delta$. In view of Eq.~(\ref{SE_1}) the coefficient $\alpha$ can be expected to vary between $1/3\pi\simeq 0.106$ and $1/2\pi\simeq 0.159$ for the stick and slip boundary condition, respectively.     

Perhaps one of the simplest and transparent variants of the SE relations for simple fluids has been derived by Zwanzig from the relations between the transport coefficients and properties of collective excitations.~\cite{ZwanzigJCP1983} His result for the self-diffusion coefficient is
\begin{equation}\label{DZ}
D=\frac{T}{3\pi}\left(\frac{3\rho}{4\pi}\right)^{1/3}\left(\frac{1}{\rho mc_l^2\tau}+\frac{2}{\rho mc_t^2\tau}\right),
\end{equation} 
where $m$ is the mass, $c_l$ and $c_t$ are the longitudinal and transverse sound velocities, related to the elastic response of fluids to high-frequency perturbations,~\cite{ZwanzigJCP1965} and $\tau$ is the relaxation time (lifetime for cell jumps in Zwanzig terminology). For a critical analysis of assumptions used in Zwanzig derivation see e.g. Ref.~\onlinecite{MohantyPRA1985}. If one further assumes that the relaxation time involved in Eq.~(\ref{DZ}) is just the Maxwellian shear relaxation time, $\tau= {\eta}/\rho m c_t^2$, we obtain the SE relation of the form
\begin{equation}\label{DZ1}
D\eta(\Delta/T)= 0.132\left(1+\frac{c_t^2}{2c_l^2}\right).
\end{equation}
Equation (\ref{DZ1}) implies that the actual value of $\alpha$ depends on the ratio of the transverse and longitudinal sound velocities.~\cite{Comment1} In some cases, this ratio can be unambiguously related to the pair interaction potential operating in fluids and then there is a certain level of predictivity regarding the value of $\alpha$. The purpose of this Research Note is to provide some related evidence. 

\begin{figure}
\includegraphics[width=8cm]{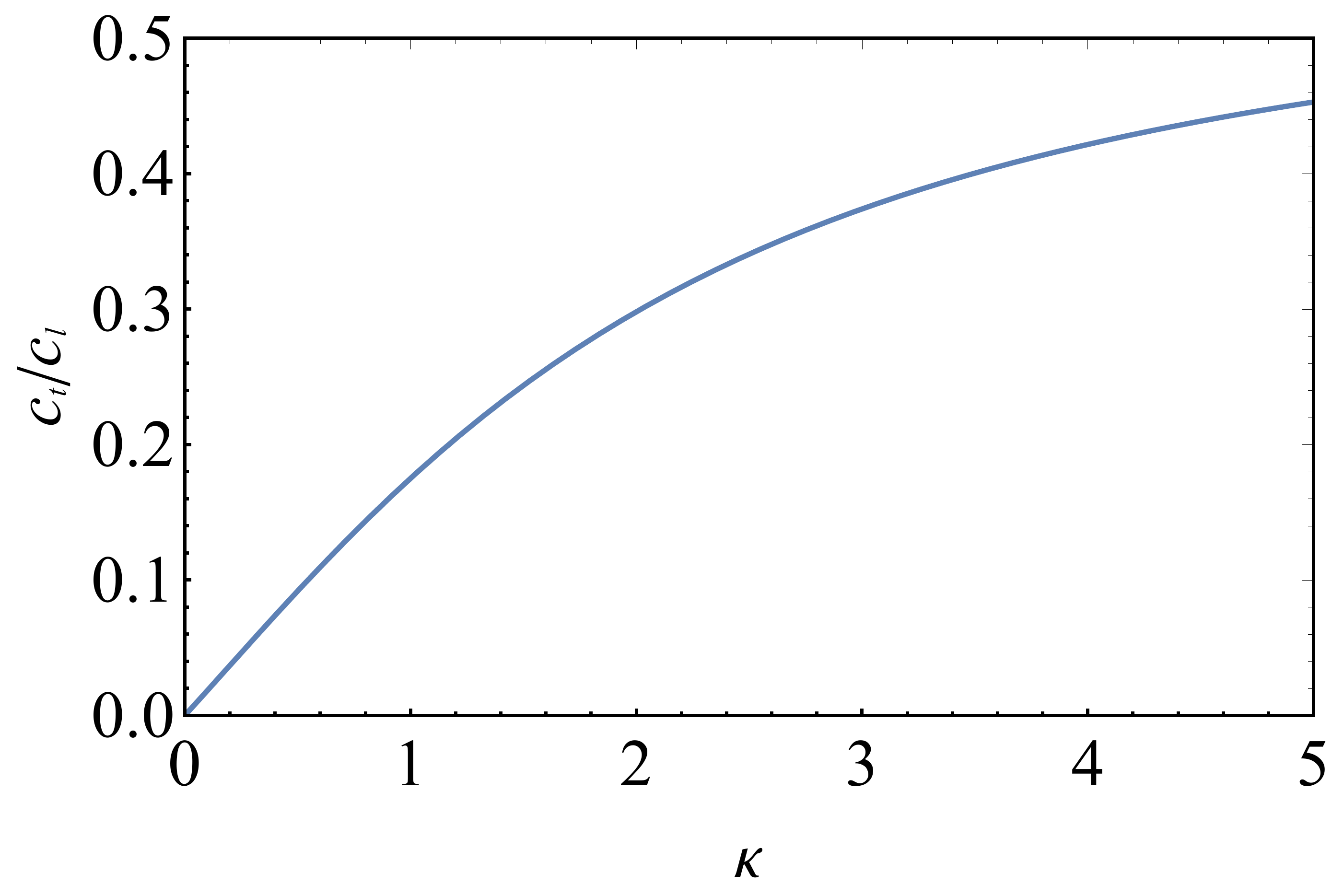}
\caption{Ratio of the transverse to the longitudinal sound velocity $ c_t/c_l$ of strongly coupled Yukawa fluids as a function of the screening parameter $\kappa=a/\lambda$. }
\label{Fig1}
\end{figure}

(i) For soft long-ranged repulsive interactions the strong inequality $c_l\gg c_t$ holds. A good example is the one-component plasma (OCP) model with Coulomb ($\propto 1/r$) interaction between the particles.~\cite{Baus1980} In this case the dispersion relation of the longitudinal mode has a non-acoustic character $\omega\simeq \omega_{\rm p}$, where $\omega_{\rm p}$ is the plasma frequency, and thus $c_t/c_l\equiv 0$. We should expect $\alpha\simeq 0.13$ in this case, which is in fair agreement with  the reported OCP result of $\alpha\simeq 0.14$, based on extensive molecular dynamics (MD) simulations.~\cite{DaligaultPRE2014} Similar situation occurs for the weakly screened Yukawa (screened Coulomb) potential, $\propto \exp(-r/\lambda)/r$, when the screening parameter is $\kappa=a/\lambda\sim \mathcal{O}(1)$. This regime of interaction is of significant interest, because it often serves as a first approximation of actual interactions in complex (dusty) plasmas and colloidal suspensions.~\cite{FortovUFN,FortovPR,KhrapakPRL2008,ChaudhuriSM,IvlevBook} The ratio of transverse-to-longitudinal sound velocity in strongly coupled Yukawa fluids is plotted in Fig.~\ref{Fig1} (this is a new calculation based on the approach from Ref.~\onlinecite{KhrapakIEEE2018}). The condition $c_l\gg c_t$  for the considered soft long-range repulsive interaction correlates very well with the coefficient $\alpha \simeq 0.126$ reported recently from the analysis of a large body of independent data on shear viscosity and self-diffusion coefficients of strongly coupled Yukawa fluids.~\cite{KhrapakJPCO2018,KhrapakAIPAdv2018} 

\begin{figure}
\includegraphics[width=8.0cm]{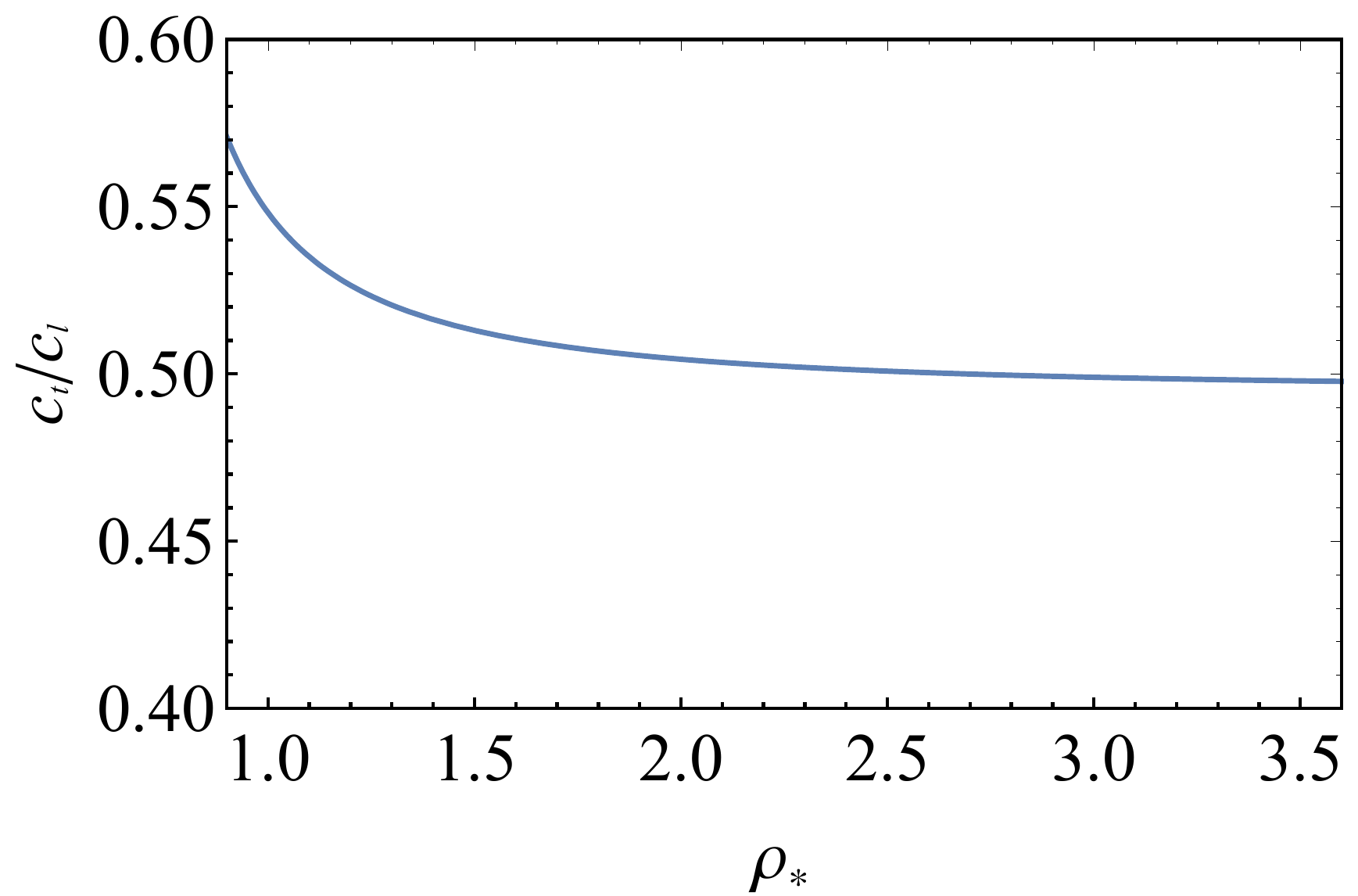}
\caption{Ratio of the transverse to the longitudinal sound velocity $c_t/c_l$ of a Lennard-Jones fluid at the fluid-solid coexistence as a function of the reduced density  $\rho_*$. }
\label{Fig2}
\end{figure}

(ii) For steeper repulsive potentials the ratio of transverse to the longitudinal sound velocity increases to $\simeq 0.5$ (see Fig.~\ref{Fig1}). For instance, for the inverse-power-law (IPL) repulsive potential, $\propto r^{-n}$, the sound velocity ratio can be roughly expressed (for $n>3$) as~\cite{KhrapakSciRep2017} 
\begin{equation}\label{IPL}
\frac{c_t}{c_l}\simeq \sqrt{\frac{n-3}{3n+1}},
\end{equation} 
where only the configurational contribution to the sound velocities (dominant at high densities) is retained. In the regime $4<n<10$, the sound velocity ratio increases from $\simeq 0.28$ to $\simeq 0.48$. The corresponding SE coefficient varies from $\simeq 0.14$ to $\simeq 0.15$. This correlates very well with MD simulations results from Ref.~\onlinecite{HeyesPCCP2007}, which are scattered around $\alpha\simeq 0.15$ in the considered softness range. 

(iii) Similar picture is characteristic for dense Lennard-Jones (LJ) liquids not too far from the fluid-solid coexistence, as Fig.~\ref{Fig2} shows (this figure reports a new calculation, the details will be given elsewhere). For the present purpose, it is important that for densities above the triple-point value, the ratio $c_t/c_l$ at freezing varies very little and has a typical magnitude of $\simeq 0.5$. Equation (\ref{DZ1}) predicts $\alpha\simeq 0.15$, which is in remarkable agreement with extensive MD simulation results demonstrating that not too far from the fluid-solid coexistence $\alpha\simeq 0.146$.~\cite{CostigliolaJCP2019}  

(iv) In the limit of very steep hard-sphere-like (HS) interaction one may expect  $c_t/c_l\simeq 1/\sqrt{3}$~\cite{BalucaniBook,ZwanzigJCP1965,KhrapakPoP2016} (see, for instance Eq.~(\ref{IPL}) above). This would lead to $\alpha\simeq 0.15$. The situation, however, is not trivial, because the standard expressions for elastic moduli and sound velocities~\cite{ZwanzigJCP1965,BalucaniBook} are diverging and should not be applied when approaching the HS limit.~\cite{KhrapakJCP2016,KhrapakSciRep2017,KhrapakNEW} Although, collective excitations in HS fluids have recently been studied by means of MD simulations,~\cite{BrykJCP2017} it is difficult to get an accurate enough estimate of $c_t/c_l$ from the data presented. A recent calculation, based on Miller's expressions for the shear and bulk moduli of HS fluids,~\cite{MillerJCP1969} has reported $c_t/c_l\simeq 0.38$ in the dense fluid regime (this indicates that the dependence of $c_t/c_l$ on the potential softness can be non-monotonous). This sound velocity ratio produces $\alpha\simeq 0.14$, which is somewhat below MD simulation results, placing $\alpha$ in the range from $1/2\pi \simeq 0.159$ to $1/6\simeq 0.167$ in the dense HS fluid regime.~\cite{OhtoriJCP2018} On the other hand, the used theoretical value of $c_t/c_l$ is consistent with the Poisson's ratio $\nu\sim 0.4$ of an fcc HS solid at a thermodinamically unstable density corresponding to fluid-solid coexistence~\cite{FrenkelPRL1987} (the ratio of the elastic sound velocities is not expected to vary much across the fluid-solid phase transition). More detailed analysis of the important HS interaction limit is clearly warranted.   

(v) Quite generally, since the  Poisson's ratio can vary between $-1$ and $1/2$ for stable isotropic materials,~\cite{Greaves2011} the sound velocity ratio has an upper limit of $\sqrt{3}/2$, resulting in the upper limit of $\alpha\simeq 0.181$ as obtained by Zwanzig.~\cite{ZwanzigJCP1983} It is not clear at this point what kind of interaction potential is required to maximize the ratio $c_t/c_l$.

\begin{table}
\caption{\label{Tab1} The coefficient $\alpha$ in the SE relation (\ref{SE_2}) of several liquid metals at the corresponding melting temperatures. Evaluated using the data tabulated in Ref.~\onlinecite{Battezzati1989}, except the diffusion coefficients of Cu and Ni, which are taken from Ref.~\onlinecite{Meyer2015}.}
\begin{ruledtabular}
\begin{tabular}{lcccccccc}
Metal & Li & Na & K & Rb & Ni &Cu & Ag &  In   \\ \hline
$\alpha$ & 0.161 & 0.142 & 0.173 & 0.178 & 0.156 & 0.165 & 0.156 & 0.160  \\
\end{tabular}
\end{ruledtabular}
\end{table}

(vi) March has pointed out the relevance of the Zwanzig result (\ref{SE_2}) for liquid metals above their melting temperature.~\cite{MarchJCP1984} The actual interaction potentials operating in various liquid metals are quite complex and seldom known accurately. However, some properties can be already explained based on soft repulsive interactions (e.g. of Yukawa type). Therefore, it makes sense to compare existing experiemental results with theoretical expectations. For liquid metals the actual ratio of sound velocities is again seldom known. Sometimes it is assumed to be around  $c_t/c_l\simeq 1/\sqrt{3}\simeq 0.58$, based on the generalized Cauchy relation.~\cite{MorkelPRE1993} Experimentally determined values for Fe, Cu, and Zn are somewhat lower, within the range $c_t/c_l\simeq 0.43-0.48$.~\cite{Hosokawa2015} All this points towards $\alpha\simeq 0.15$ for liquid metals. The coefficient $\alpha$ at the melting temperature has been re-evaluated here for those liquid metals, for which simultaneous data on diffusion and viscosity coefficients are available. For some elements, in particular for alkali metals, the experimental values of $\alpha$ are reasonably close to the above expectation, as documented in Table~\ref{Tab1}. There are, however, liquid elements for which experimental values of $\alpha$ are significantly higher and even exceed the maximum possible value of the Zwanzig model. For example, $\alpha\simeq 0.209$ (Zn), 0.192 (Hg), 0.219 (Ga), 0.187 (Sn), and 0.224 (Pb), according to the data tabulated in Ref.~\onlinecite{Battezzati1989}. Whether there are some physical explanations behind these deviations, or this merely reflects the quality of the data presently available needs to be clarified. 

(vii) It is known that self-diffusion coefficients obtained from MD simulations may depend on the simulation cell size. A correction of the form
\begin{equation}\label{Dinf}
D_{\infty}\simeq D_N+2.84 \frac{T}{6\pi \eta L},
\end{equation}
has been suggested in the literature.~\cite{DunwegJCP1993,YehJPCB2004} Here $D_{\infty}$ is the infinite-size system diffusion coefficient, $D_N$ is the diffusion coefficient evaluated from the simulation of $N$ particles in a cubic cell of size $L$ with periodic boundary conditions. This correction has demonstrated reasonable quantitative accuracy when applied to simulations of water, LJ fluids, and HS fluids. In contrast to diffusion, the shear viscosity does not exhibit pronounced system-size dependence for these systems.~\cite{YehJPCB2004} Two related comments are appropriate here. First, some of the MD data discussed above are corrected for the finite system size (e.g. those from Ref.~\onlinecite{OhtoriJCP2018}), but some are apparently not. Such a correction would somewhat increase the actual value of the SE coefficient. Second, using the identity $\rho L^3= (L/\Delta)^3=N$ and assuming an ``average'' value of SE coefficient $\alpha=0.15$  (that is $D_N\eta \Delta/T\simeq 0.15$) we immediately get from Eq.~(\ref{Dinf}):
\begin{equation}\label{Dinf1}
D_{\infty}\simeq D_N(1+N^{-1/3}).
\end{equation}        
This particularly simple expression for system-size corrections can be useful when results from different simulations are compared. Equation (\ref{Dinf1}) indicates that for a system of $N=1000$ particles the MD-based self-diffusion coefficient can be $\simeq 10\%$ lower than that in a similar infinite system.

To summarize, the Zwanzig's approximate theory for the relation between the coefficients of self-diffusion and viscosity of fluids has been re-examined. The advantage of Zwanzig's formulation, compared to the original SE formula, is that it does not require the choice of an effective hydrodynamic diameter and of the appropriate boundary condition (e.g., stick or slip). For simple fluids, the coefficient in SE relation can be expressed using the ratio between the transverse and longitudinal sound velocities. Within the Zwanzig's theory the SE coefficient $\alpha$ is bounded by $\simeq 0.13$ from below and $\simeq 0.18$ from above. It has been demonstrated here that for soft repulsive interactions the value of the SE coefficient tends to its lower boundary. In other cases considered, including LJ and HS interactions, the theoretical value of $\alpha$ is close to $\simeq 0.15$. For many liquid metals the experimental value of the SE coefficient falls in the expected range. There are, however, some metal elements for which $\alpha$ is considerable higher. One of the potential further developments is to understand whether this discrepancy is due to a more complex physics involved or just reflects the accuracy of the experimental data presently available.               
               
I thank Andreas Meyer for reading the manuscript.

\bibliographystyle{aipnum4-1}
\bibliography{SE_References}

\end{document}